\documentstyle[epsf]{article}
\begin{document}
\renewcommand{\thefootnote}{\fnsymbol{footnote}}
\begin{flushright}
KEK-preprint-93-3\\
\end{flushright}
\vskip -2cm
\epsfysize3cm
\epsfbox{kekm.epsf}
\begin{center}
{\bf \Large
Feasibility Study of Single-Photon Counting
Using a  Fine-mesh Phototube for an Aerogel Readout
\footnote{publushed in Nucl. Instr. Method {\bf A 332} (1993) 129.}}\\
\vskip 1cm
R. Enomoto\footnote{internet address: enomoto@kekvax.kek.jp.},
T. Sumiyoshi, K. Hayashi, I. Adachi\\
{\it National Laboratory for High Energy Physics, KEK,
1-1 Oho, Tsukuba-city, Ibaraki, 305 Japan}\\
\vskip 0.5cm
S. Suzuki, H. Suzuki\\
{\it Hamamatsu Photonics K. K.,
314-5, Shimokanzo, Toyooka Vill., Iwata-gun, Shizuoka, 438-01
Japan}\\
\end{center}
\begin{abstract}
The fine-mesh phototube is one type of photodetector which can be
used under a strong magnetic field.
For an aerogel readout, the single-photon
detection efficiency should be close to 100\% in order to identify
particle species. We carried out a feasibility study of
single-photon counting using fine-mesh phototubes, and obtained a
possible solution.
\end{abstract}

\section{Introduction}

Experimental high-energy physics in recent years has seen
an increased demand for highly efficient photodetectors that can
operate under a strong magnetic field.
This is particularly true for the read-out of aerogel Cerenkov
counters (Aerogel) \cite{refaero} and ring-imaging
Cerenkov detectors (RICH) \cite{refrich}.
Among other photo-detection devices, such as microchannel plate
phototubes (MCP)
and solid state detectors at low temperature, fine-mesh
phototubes (FM) \cite{reffm} seem to be the most promising due to
their detection area, high quantum efficiency, radiation hardness,
and stability.
An FM phototube may also be used as the basis of a position-sensitive
photon detection device by using segmented anodes (multi-anode
phototube, MAPT) \cite{refmapt}.
In this paper we present our feasibility study of FM phototubes used
for single-photon detection. Our main interest is in its application to a
charged-particle identification system based on aerogel Cerenkov counters.
The results of our simulation and experimental studies concerning this area
are presented.

\section{Requirement for an Aerogel Readout}

The motivation of our study was to develop a charged-particle
($\pi/K$) identification system for a future B-factory
experiment \cite{refb}.
In an $e^+e^-$ collision at $\Upsilon_{4s}$, a good $\pi/K$
separation capability up to momenta of 3-4 GeV is required.
Threshold-type aerogel Cerenkov counters
(reflective index: n=1.02 - 1.006) can be used for this
purpose \cite{refslac}.
In this case, the typical number of Cerenkov photons
produced by a relativistic
pion above the
Cerenkov threshold is not very large, about 4-5 \cite{refaero}.
A kaon is identified by detecting a charged particle associated
with no corresponding Cerenkov photons. It is thus critical to
employ a photon detection device with a high single-photon
detection efficiency. Another important consideration is that the
detector system must operate in a strong magnetic field,
typically 1-Tesla, which is used for charged-particle momentum
analysis in the experimental facility \cite{refb}.

\section{Fine-mesh Phototube}

\subsection{General}

The fine-mesh (FM) phototube which we tested was developed by
Hamamatsu Photonics K. K. (R2490-05, assembly-type H2611) \cite{reffm}.
Its outer diameter is 2 inches and the sensitive area comprises
a 36-mm$\phi$ bialkali photocathode.
The number of fine-mesh dynode stages is 16.
The gap between neighboring dynodes is approximately 1.0 mm;
that between the photocathode and
the first dynodes is less than 3 mm.
A typical gain is $\sim10^{7}$ ( at 2500V) without any magnetic field.
Under a 1-Tesla axial field, the gain drops by 1/30 \cite{reffm}.
The phototube can
be used within an inclination angle of
 45 degree with respect to the magnetic field axis \cite{reffm}.
This phototube also shows a good transit time spread of 400
psec \cite{reffma}.

\subsection{Fine-mesh Dynodes}

\begin{figure}
\epsfysize2.5cm
\epsfbox{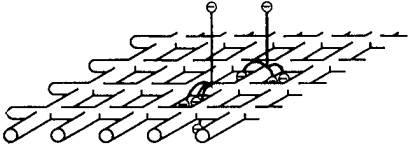}
\caption{Basic configuration of a fine-mesh dynode.}
\label{photofm}
\end{figure}

A basic configuration of the fine-mesh
is shown in Fig \ref{photofm}.
In total, the opacity
is optically approximately 50\%.
The photocathode and dynodes are placed perpendicular
to the phototube axis,
so that the electrostatic field is almost axial. We therefore can not
expect a much higher hit probability at the first dynode than
the optical opacity.
It is quite
difficult to improve the hit probability of photoelectrons at the first
dynode using this kind of phototube.

\section{Single Photon Counting by H2611}

\begin{figure}
\epsfysize7cm
\epsfbox{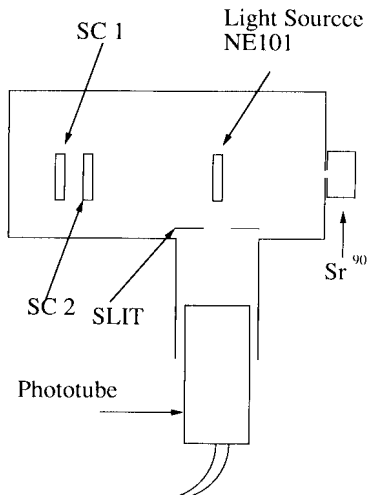}
\caption{Experimental setup for single-photon counting.}
\label{expsetup}
\end{figure}

Fig \ref{expsetup} shows a schematic view of the setup of our single-photon
counting experiment.
The incident photons are created by a thin (500$\mu$mt) scintillator
(NE101) illuminated by a $\beta$-ray source ($Sr^{90}$).
The trigger signal
is created by a coincidence signal from
two scintillators located downstream of this light source.
To control the photon yield, approximately one photon per trigger, a
collimator is used between the light source scintillator and the
phototube to be tested.
The pulse height
of the phototube signal is digitized by a LeCroy 2249W ADC
\cite{reflecroy}.

\begin{figure}
\vskip -1cm
\epsfysize9cm
\epsfbox{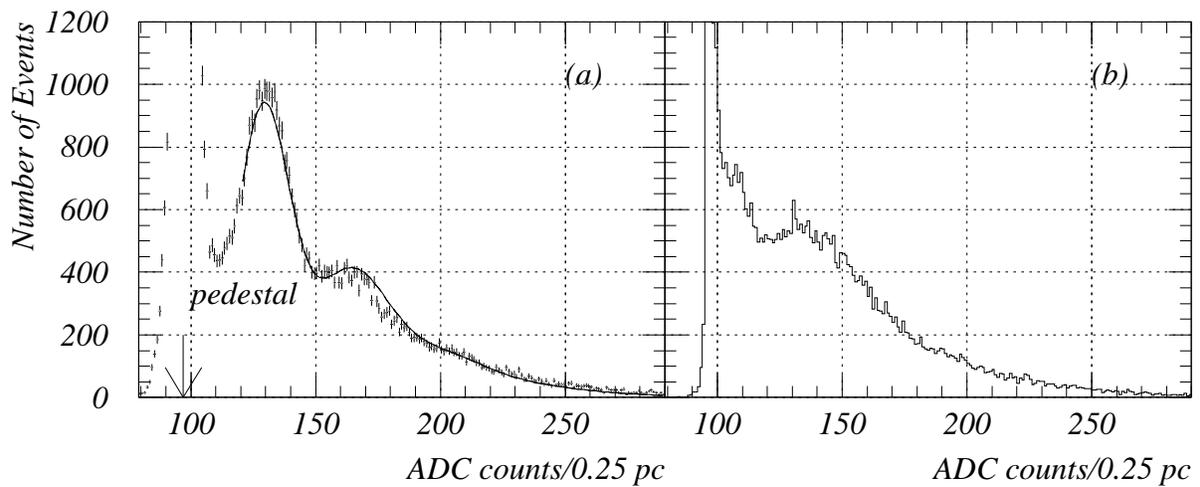}
\vskip -1cm
\caption{Pulse-height spectra obtained by the test:
(a) for R3241, and (b) for H2611 (FM phototube).
In (a), a best-fitted Poisson function is plotted (curve).
The pedestals
for these plots were 97 ch.}
\label{expsingle}
\end{figure}

A calibration was carried out using the R3241
(Hamamatsu Photonics; GaAs-P first dynode);
the measured pulse-height
spectrum is shown in Fig \ref{expsingle}-(a).
A fit with a Poisson distribution gave the mean photoelectron yield:
$\mu$=1.2 per event.
After this calibration, we carried out a test using
the H2611 (Assembly of R2490-05: FM phototube).
The pulse-height spectrum obtained by this FM phototube is shown in Fig
\ref{expsingle}-(b). The high voltage was adjusted to have a total gain of
$6\times 10^{7}$. Thus, the expected ADC count for a single photon
was considered to be 137 channel (ch.), where the pedestal was 97 ch.
Although a clear dip structure at around 120-ADC counts was seen
in the data, we could not obtain a clear peak that was consistent with
the Poisson distribution. In the following section we examine
the observed signal pulse-hight structure.

\section{Simulation Study}

\subsection{Simple Simulation}

\begin{figure}
\vskip -1cm
\epsfysize9cm
\epsfbox{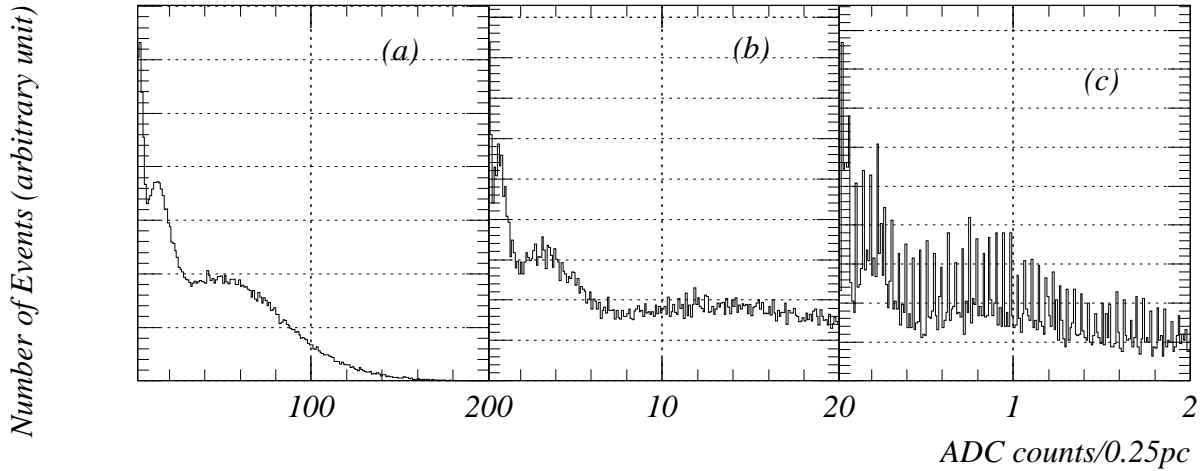}
\vskip -1cm
\caption{Pulse-height distribution obtained by a simulation of
single-photoelectron events: (a) ADC range 0-200, (b) ADC range 0-20, and
(c) ADC range 0-2.}
\label{mcsingle}
\end{figure}

At first, we started by making a simple assumption:
that the hit probability of electrons on
each dynode was 60\%.
The total gain of phototube was assumed to be $2.5\times10^7$
by 12 dynodes, while each dynode gain was assumed to be equal.
A single photoelectron
was generated at the photocathode and the number of secondary
electrons at each dynode was randomly created
according to Poisson distributions.
The resulting pulse-height distributions are shown
in Figs \ref{mcsingle}-(a),(b),
and (c) with three different horizontal bin sizes.
Accidentally, the pulse-height distribution of Fig \ref{mcsingle}-(a)
coincided
with Fig \ref{expsingle}-(b) by the FM phototube.
It is shown that single-photon events would not emerge as a clear peak
but, rather, they create a fractal-structure in the pulse height
spectrum.

A fractal structure appears when a process is described
in terms of the function
$x^\alpha$ \cite{reffractal}.
In the FM phototube case, $x$ is the dynode gain and $\alpha$
is the effective number of dynodes, i.e., the dynode hit probability times
the number of dynodes; $\alpha$ is typically a noninteger.
The pulse-height distribution should therefore have a similar shape in any
ADC range; it is thus impossible to obtain the peak structure.
The peak around 130-ADC counts
 shown in Fig \ref{mcsingle}-(a) is considered to be the case
in which
a photoelectron hits the first dynode; the peak around
110-ADC count is for
a photoelectron hitting the second dynode.
The pulse-height difference between two peaks is the same as the dynode
gain, and the ratio of each peak entry is same as the dynode hit probability.
Thus, it is again understood that a single-photon peak cannot
be observed by FM phototubes,
even by increasing the first dynode gain or
by improving the first dynode hit probability.
Even upon setting a low threshold,
an inefficiency remains.

\subsection{Possibility of Single-Photon Counting}

In order to cure this situation,
\begin{enumerate}
\item obtaining a higher gain, and/or
\item making the pulse height distribution harder
\end{enumerate}

\noindent
are necessary. Both situations can be realized by increasing the number of
dynodes. By this method, each dynode gain can be reduced and the gap between
peaks can be made smaller. Events
which started multiplication at deeper dynodes can thus be saved.

Using the same simulation we increased the number of dynodes to
15 and 20.
The pulse-height distributions
became
harder. In addition the waving
structure disappeared due to the low gain of each dynode.

In order to make a single-photon peak, we must
significantly increase the probability of the
first dynode hit.
We checked this using the same simulation and
concluded that a probability greater
than 90\% is necessary. If
we provide such a high opacity in fine-mesh dynodes, secondary
electrons cannot be transported to the second dynode. Therefore, we may
need other kinds of dynode structures, for example wire dynodes (grids) with
an alternating high voltages to produce a highly focused electric field.

\section{Improvement of the Fine-mesh Phototube}

According to the results obtained by the simulation, we developed
phototubes with a larger number of dynodes, i.e., 19-dynodes (H2611(19)) and
24 dynodes (H2611(24)). The former one was made using the same
type of glass tube
as that used for the
R2490-05 phototube; for the latter, a longer glass tube was
made.

The typical gains at 2500V
obtained by these phototubes without any magnetic field were
$6\times10^{7}$ and $1.2\times10^{9}$ for H2611(19) and H2611(24),
respectively.
There was no increases in the dark current
due to these improvements.

\section{Experimental Study}

\subsection{Experimental Results}

\begin{figure}
\vskip -1cm
\epsfysize9cm
\epsfbox{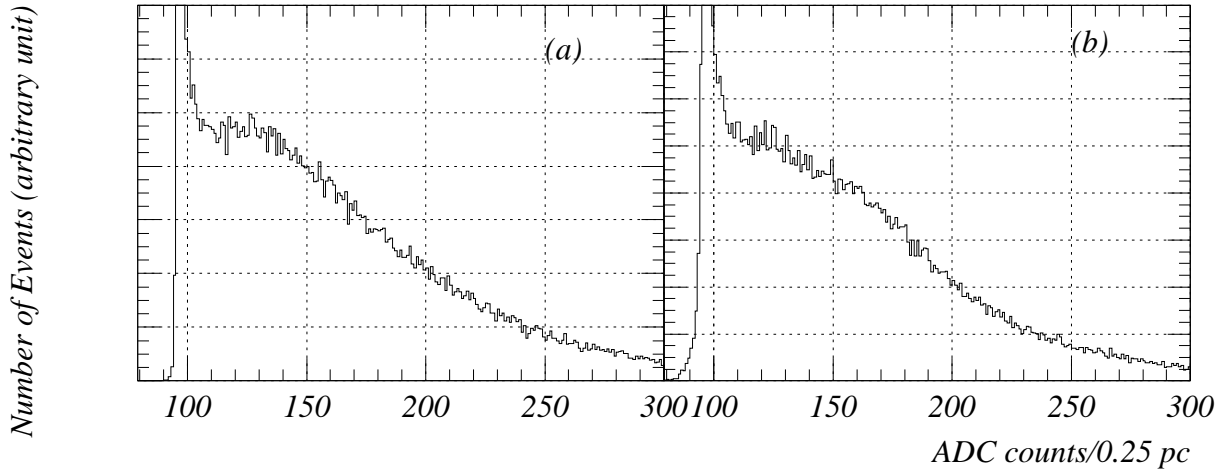}
\vskip -1cm
\caption{Single-photon spectra obtained by the improved phototubes:
(a) H2611(19) and H2611(24).}
\label{expnd}
\end{figure}

We carried out tests using the setup shown in Fig \ref{expsetup}.
The single-photon spectra obtained by these two phototubes are shown in Figs
\ref{expnd}-(a) and (b) for H2611(19) and H2611(24), respectively.
The high voltages were adjusted so as to obtain gains of $6\times10^{7}$
in both cases.
The single-photon efficiencies were consistent with
that of the H2611 (Fig \ref{expsingle})
at the extrapolation to a zero-threshold.
The waving structure disappeared and the spectrum shapes became harder,
as expected, due to the previously described simulation.

\begin{figure}
\vskip -2cm
\epsfysize10cm
\epsfbox{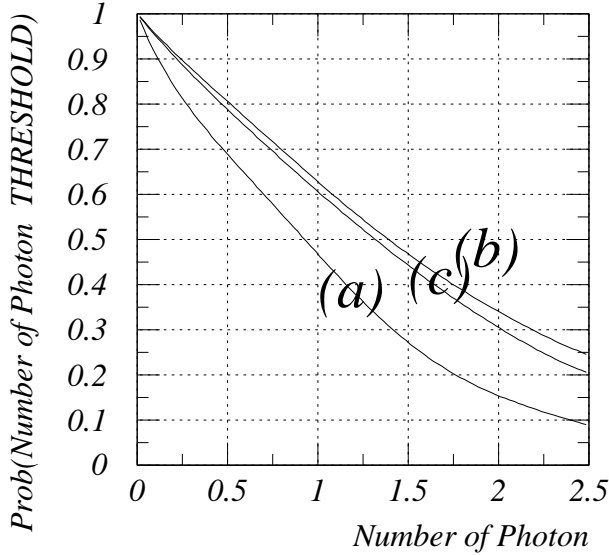}
\vskip -1cm
\caption{Single-photon efficiencies versus threshold values:
(a) for H2611, (b) for H2611(19), and H2611(24).}
\label{expeff}
\end{figure}

We tried to obtain relations of
the single-photon efficiencies versus
the threshold values by the following method.
We derived the relationship between the integrated contents,
 which were greater
than the threshold values, versus the thresholds. The normalizations were
carried out
using polynomial-fits to these relationships at zero-threshold points.
The results for three kinds of phototubes are given in Fig \ref{expeff}.
There was a clear improvement of H2611(19) compared to
that of H2611. The results of H2611(19)
and H2611(24) were similar to each other within the experimental errors
(including the statistical and systematic errors).
In order to obtain the single-photon efficiency greater than 90\%,
we must lower the thresholds to 0.1-photon for the H2611 and 0.25-photon
for the improved phototubes; this was quite easy for the improved tubes
owing to the high gain. In case of the H2611, it was suggested that use of
a preamplifier under low-gain operation would yield a high efficiency.
Using the simulation described so far, the suggested gain was
less than $10^6$.

\subsection{Test under Magnetic Field}

\begin{figure}
\vskip -2cm
\epsfysize10cm
\epsfbox{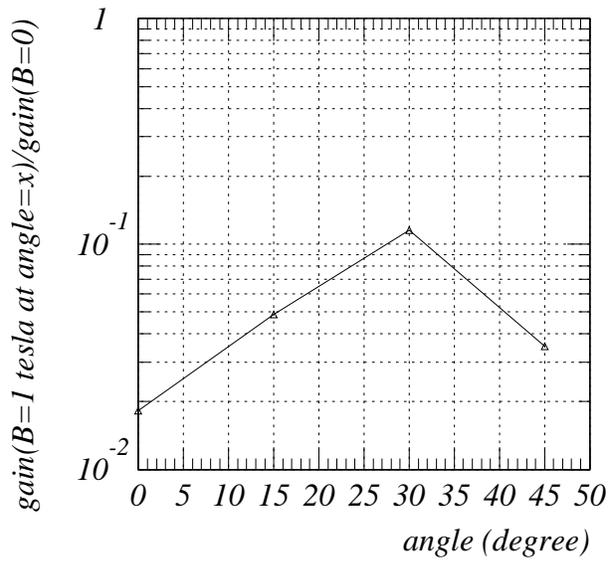}
\vskip -1cm
\caption{Gain under a 1 tesla magnetic field. The horizontal
axis is the inclination angle with respect to the magnetic field
axis. The vertical scale is the normalized gain with respect to the
gain without a magnetic field.}
\label{bangle}
\end{figure}

The gain and single-photon measurements were carried out under a 1 tesla
magnetic field. The phototube used in these tests was
the H2611(24). Fig \ref{bangle}
shows the angle dependence of the phototube gain.
The vertical scale is the gain at an inclination angle with respect to the
magnetic field axis normalized by the gain at B=0.
The angle dependence of the gain was similar to the case of the H2611(16),
except for the overall scale factor. The gain drop observed for the H2611(24)
was about 1/50 at 0 degree,
larger than that of the H2611(16) ($\sim$1/30).
It is therefore hard to read it out without using a preamplifier.
The high-voltage dependence
of the gain was also measured. The gain was typically described by the
function $V^{\alpha}$; we could not observe any magnetic field
dependence of $\alpha$.

\begin{figure}
\vskip -2cm
\epsfysize10cm
\epsfbox{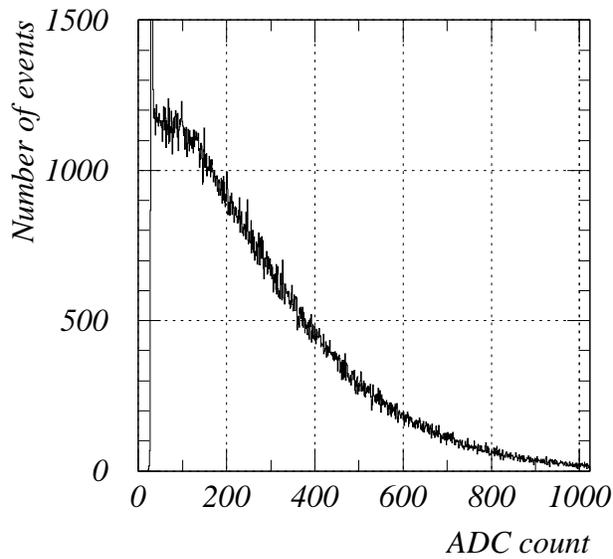}
\vskip -1cm
\caption{Single-photon spectrum by the H2611(24) at 1 tesla with
an inclination angle of 15 degrees.}
\label{bsingle}
\end{figure}

The single-photon spectrum was measured at 1 tesla with an inclination angle
of 15 degrees (Fig \ref{bsingle}).
In this case we used a preamplifier because
of the low gain. The light source was a pico-second laser; a histogram was
obtained by self-triggering. The histogram shown in Fig \ref{bsingle}
is a background-subtracted type (i.e. on-source minus off-source run).
The single-photon spectrum was similar to Fig \ref{expnd} (b). Therefore,
single-photon detection under a magnetic field was proven.

\section{Summary}

We carried out systematic studies on single-photon counting using fine-mesh
phototubes, especially for the purpose of aerogel readouts. The studies
were carried out using both simulations and experiments.
We found one possible solution to improve the single-photon
efficiency: to increase the number of fine-mesh dynodes.
We concluded that fine-mesh phototubes can be used for an aerogel Cerenkov
counter readout at a future B-factory.

\section*{Acknowledgement}
We appreciate the support given by Prof. H. Sugawara, Prof. S. Iwata,
Prof. M. Kobayashi, and
Prof. F. Takasaki in carrying out this R and D.
We also thank Prof. F. Takasaki and Prof N. Toge for
their valuable discussions.
We greatly appreciate the technical staff of KEK and Hamamatsu
Photonics K. K.


\begin{thebibliography}{99}
\bibitem{refaero}
A. Onuchin et al., to be published in Nucl. Instr. Meth. {\bf A 315}
(1992) 517.
\bibitem{refrich}
J. Seguinot, CERN-EP/89-02, July (1989).
\bibitem{reffm}
A. Sawaki et al., IEEE Trans. Nucl. Sci., {\bf NS-31, 1} (1984) 442;
M. D. Rousseau et al., IEEE Trans. Nucl. Sci., {\bf NS-30, 1} (1983) 479;
H. Kume et al., IEEE Trans. Nucl. Sci., {\bf NS-32, 1} (1985) 355;
S. Suzuki et al., IEEE Trans. Nucl. Sci., {\bf NS-33, 1} (1986) 377.
\bibitem{refmapt}
F. Takasaki et al., Nucl. Instr. Meth., {\bf A 260} (1987) 447.
\bibitem{refb}
Proceedings of ``Physics and detectors for KEK asymmetric
B factory", KEK Proc 91-07, KEK, Tsukuba, Japan (1991).
\bibitem{refslac}
P. A. Coyle et al., SLAC-PUB-5594, Sep. (1991).
\bibitem{reffma}
B. Lekovar et al., Nucl. Instr. Meth., {\bf A123} (1975) 145;
F. Takasaki et a., Nucl. Instr. Meth., {\bf A228} (1985) 369.
\bibitem{reflecroy}
See Nuclear Products Summary, LeCroy.
\bibitem{reffractal}
H. Takayasu, ``Fractal in the physical science",
Manchester Univ. Press (1990) 157.
\end{thebibliography}
\end{document}